\documentclass[journal=jacsat,manuscript=article]{achemso}
\usepackage[version=3]{mhchem} 
\usepackage{xcolor}
\usepackage{soul}

\newcommand{\mrm}[1]{\mathrm{#1}}
\newcommand{\mbf}[1]{\mathbf{#1}}

\newcommand{\ave}[1]{\left\langle{#1}\right\rangle}

\newcommand{\mcal}[1]{\mathcal{#1}}

\author{Emily Krucker-Velasquez}
\email{ekrucker@mit.edu}
\affiliation[ChemE]{Department of Chemical Engineering, Massachusetts Institute of Technology, Cambridge, Massachusetts 02139, USA}

\author{Martin Z. Bazant}
\affiliation[ChemE]{Department of Chemical Engineering, Massachusetts Institute of Technology, Cambridge, Massachusetts 02139, USA}

\author{Alfredo Alexander-Katz}
\affiliation[DMSE]{Department of Materials Science and Engineering, Massachusetts Institute of Technology, Cambridge, Massachusetts 02139, USA}

\author{James W. Swan}
\affiliation[ChemE]{Department of Chemical Engineering, Massachusetts Institute of Technology, Cambridge, Massachusetts 02139, USA}

\title[PMF]{Potential of mean force and underscreening of polarizable colloids in concentrated electrolytes}

\begin{document}







\begin{abstract}
This study uses advanced numerical methods to estimate the mean force potential (PMF) between charged, polarizable colloidal particles in dense electrolytes. We observe that when the Debye screening length, $\lambda_{\mathrm{D}}$, is below the hydrated ion size, the PMF shows discernible oscillations of purely electrostatic origin as opposed to chemical affinity, in addition to the expected decay in DLVO theory. Moreover, our findings suggest concentrated electrolytes significantly less efficient at muting electrostatic interactions in electrostatically stabilized colloidal suspensions, potentially having significant implications for our understanding of colloidal stability and the forces that govern the behavior of concentrated charged soft matter systems beyond DLVO theory.
\end{abstract}
\section{Introduction}
\indent Colloidal suspensions are ubiquitous in nature. In modern society, their utility spans from environmental applications such as water treatment\cite{li_potential_2022}, to critical roles in energy delivery and storage\cite{che_highenergydensity_2023}, to healthcare and nanotechnology\cite{gupta_interfacing_2023}. In emerging fields of nanotechnology and biomedicine, metallic nanoparticles (MNPs) are versatile and powerful tools. Their distinctive electronic and optical properties, coupled with their nanoscale dimensions, enable a myriad of applications. For example, silver and gold nanoparticles are used for bioimaging.\cite{si_gold_2021,hang_plasmonic_2024}  In addition, NPMs surfaces can be meticulously engineered to bind to specific biomolecules and/or bioactive materials, allowing for precise targeting in drug delivery systems\cite{turrina_application_2023,hang_plasmonic_2024} and detection of biological markers\cite{kaefer_implantable_2021} and chemical analytes\cite{montes-garcia_chemical_2021} with exceptional sensitivity. The widespread application of colloidal suspensions hinges on our ability to precisely manipulate the interparticle potential between particles. This control over the interactions at the microscopic level is key to determining not only the stability but also the dynamic response\cite{steimel_artificial_2014} of these suspensions.\\
\indent Our understanding of electrostatically stabilized colloidal suspensions is primarily framed within the context of the DLVO theory (Derjaguin-Landau-Verwey-Overbeek), a cornerstone of colloid science and surface chemistry.\cite{Israelachvili1991}  The DLVO theory proposes the superposition of two critical forces: attractive van der Waals forces and repulsive electrostatic forces\cite{Israelachvili1991}. The repulsive portion of the DLVO potential follows from Debye-Huckel (DH) theory. DH theory predicts that the repulsive forces between charged particles in an electrolyte decrease exponentially with increasing center-to-center distance between particles, $r$. This exponential decay of the electrostatic repulsion is controlled by the DH screening length ${\lambda_{\mathrm{D}}= \sqrt{ \varepsilon_{\mathrm{f}} k_{\mathrm{B}}T  /(e^2 \sum_{a=1}^M n_\nu z_\nu^2 ) }}$, or its corresponding wave vector, $\kappa_{\mathrm{D}} = \lambda_{ \mathrm{D }}^{-1}$; where ${n_{\nu}}$ and ${z_{\nu}}$ are the bulk number density and valence of species ${\nu}$ respectively, ${e}$ is the fundamental charge, ${\varepsilon_{\mathrm{f}}}$ is the solvent permittivity and ${k_{\mathrm{B}}T}$ is the thermal energy.\\
\indent Classical mean-field theories, including DLVO theory, have been instrumental in advancing our understanding of electrostatically stabilized colloidal systems. However, due to mathematical complexity, these theories often overlook important physical phenomena, such as the inability of particle's to overlap and ion-ion couplings. While proven extremely useful, classical mean-field approaches become increasingly inadequate under conditions where the system deviates significantly from the idealized scenarios\cite{Israelachvili1991}. This is the case for concentrated electrolytes.\\
\indent Recent experimental findings have sparked a reevaluation of the traditional boundaries within which DH theory is considered applicable. Lee et al\cite{Lee2017} leveraged surface force apparatus to infer the effective screening length from the force acted upon charged mica surfaces. The authors observed that when the Debye screening length is on the order of magnitude of the diameter of the hydrated ion, $2a$, the measured screening length, is found to be significantly larger than what the DH theory would predict and can be characterized by the scaling relation $\left( \lambda^{\dagger} / \lambda_{\mrm{D}} \right)\sim (2a/\lambda_{\mrm{D}})^n$ with the scaling exponent $n=3$ regardless of the identity of the ions in $1:1$ salts. This phenomenon has been referred to as \emph{anomalous} ``underscreening'' and challenges previously predicted concentration dependence by classical models (with the scaling exponent being $n=1$) and modern mean-field asymptotic approximations\cite{Andelman2019, de_souza_interfacial_2020} (with the scaling exponent being $n=2$). \\
\indent In our previous research,\cite{krucker-velasquez_2021} we studied the importance of excluded volume interactions when modeling electrolytes using spectral methods to determine the correlation length of symmetric electrolytes in bulk as a function of the hydrated radius of the ions. Our studies show significant deviations from the behavior of the ideal solution at high concentrations (approximately concentrations above $1\mrm{M}$ of $\mrm{NaCl}$) and an excess of ionic groups. We found that the correlation length falls on a master curve \emph{only when scaled on the hydrated and not the electrostatic radius}, thus underscoring the importance of excluded volume interactions commonly ignored in analytical studies due to the mathematical complexity involved. When rescaled on the hydrodynamic radius, we find $n\approx2$. This appears to be in accordance with the scales found analytically.\cite{Andelman2019, jager_screening_2023} However, when this asymptotic analysis is performed for physically feasible values of $\lambda_{\mrm{D}}$ ($ \mcal{O}(10^{-1}) -  \mcal{O}(1)$ ), $n$ is not $n=2$ but rather $n\approx1.6$. Only when $\lambda_{\mrm{D}}/a \rightarrow \mcal{O}(10^{-3})  $, the quadratic exponent, $n=2$, is reached. Moreover, the presence of ionic clusters has been associated with an effective decrease in the screening of electrostatic interactions between charged nanoparticles.\cite{li_strong_2017}\\
\indent This effective screening length controls the interparticle forces that govern colloidal stability in a wide variety of colloidal suspensions. Thus, the possibility of deviations from DH theory in the highly concentrated electrolyte regime could have significant implications for our understanding of colloidal stability and might help explain recent open questions in charged soft-matter systems such as salt-dependent phase separation reentry in weak polyelectrolytes\cite{li_salt-dependent_2023} and the salt-dependent swelling and deswelling of DNA-coated nanoparticles.\cite{reinertsen_reexpansion_2024}\\
\indent In addition to the electrolyte-mediated forces, the nature of the colloidal particles plays a crucial role in determining the interactions of the colloids with their surrounding ionic environments. Take, for instance, dielectric versus metallic particles: the former has been found to be less efficient in muting their interactions, whereas the latter showcases a greater capacity for neutralizing charges and has a huge influence on the crystallization path and structure of charged nanoparticles.\cite{coropceanu_self-assembly_2022}\\
\indent We hypothesize that important deviations from interparticle forces predicted by classical mean-field theories can also arise in metallic nanoparticles suspended in concentrated electrolytes, and this study will aid in understanding how these deviations manifest in different conditions and lengthscales, a question currently at the forefront of scientific discourse and inquiry.\cite{li_salt-dependent_2023,reinertsen_reexpansion_2024, yuan_colloidal_2022} \\
\indent To this end, we leverage enhanced sampling techniques to systematically determine the potential of mean force (PMF) between two charged and ideally polarizable colloidal particles in an explicit electrolyte. By ``ideally polarizable,'' we refer to colloids that behave as equipotential metallic bodies, with surface charge distributions fully determined by induced polarization effects. \\
\indent We employ large-scale Brownian dynamics simulations coupled with Poisson's equation and a novel immersed boundary method to accurately represent the polarizability of the colloidal particles, which has been validated for its effectiveness in capturing electrostatic interactions across dissimilar length scales.\cite{krucker-velasquez_immersed_2024} In our simulations, all charged species are modeled explicitly and the solvent is treated implicitly. To ensure precision in the computation of long-range electrostatic forces, we utilize a spectrally accurate Ewald summation method.\cite{Lindbo2011,krucker-velasquez_immersed_2024} \\
\indent The computation of the PMF for these systems is inherently demanding, primarily due to the necessity of capturing interparticle forces at long-range distances, which requires very large simulation boxes. Additionally, polarization effects spanning multiple length scales pose numerical challenges, often demanding a large number of iterations for convergence. Moreover, constructing the PMF at a single electrolyte concentration necessitates multiple independent simulations, and extending this to resolve the concentration-dependent behavior involves hundreds of simulations. These computational challenges highlight the critical importance of robust and efficient numerical methods, as they directly impact the feasibility of such an exhaustive study. To the best of our knowledge, this work represents the first comprehensive investigation of the PMF for metallic polarizable particles in explicit electrolytes across a range of concentrations addressing polarization effects.\\

\begin{figure*}[ht!]
\includegraphics[width=0.95\linewidth]{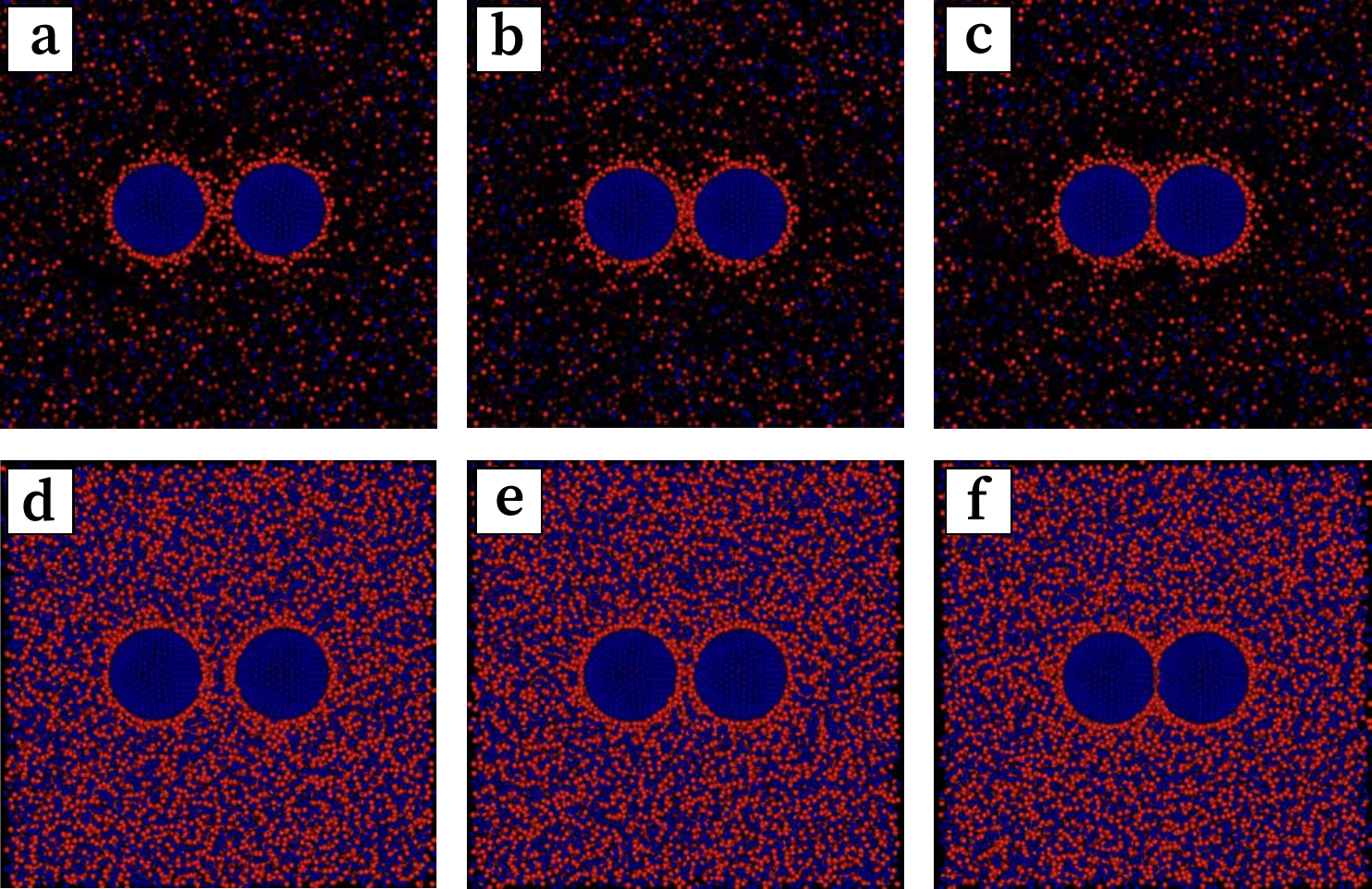} 
\caption{ Snapshots of simulations at different biasing potential resting distances, $r_0$ . Figures (a), (b), and (c), show the case for $\phi_i=0.01$ and $\epsilon=0.5 $. Panels (d), (e), and (f)}, show the case for $\phi_i=0.20$, and $\epsilon=0.5$. For all subpanels, the colloid charge bead density in units of the ion charge is $\sigma = 1.0$ and the colloid's radii $a_p\approx 15 a$.
\label{fig:Snapshots}
\end{figure*}
\section{Results and Discussion}
\indent The potential of mean force (PMF) of two particles in solution is defined as the interaction potential between these two particles \emph{including solvent interactions and effects} as a function of their separation.\cite{Macquarrie2000} The value of the PMF at a (discretized) distance $r_j$, $\mrm{PMF}_j$, is intrinsically related to the likelihood of finding these particles at this specific separation. This relationship is encapsulated in the following equation:\cite{Macquarrie2000}
\begin{equation}
    \mathrm{PMF}_j = -k_{\mrm{B}}b T \log \left \langle \chi_j\right \rangle,
\label{eq:PMF}
\end{equation}
where the term $\ave{\chi_j }$ is the expectation value of the indicator function $\chi_j$. This indicator function takes the value of one when the particles are found within the $j^{\mrm{th}}$ bin of separation and zero otherwise. \\
\indent Our model includes ions and counterions associated to the charged colloidal particles, each represented as hard, non-overlapping spheres of radius  $a$  and charge $q_i$. The colloidal particles themselves are characterized by a radius $a_p$ and a different charge $Q$. The surface of these colloidal particles is discretized by positioning $N_p$ beads, each identical in size to the ions, across the surface of the colloids as elaborated in detail in ref.\cite{krucker-velasquez_immersed_2024}. A brief explanation of the underlying assumptions in our simulations can be found in Sec. \ref{sec:Methods} and additional simulation parameters are discussed in the SI.\\
\indent The strength of the ion-ion interactions is controlled by the quantity ${\epsilon}$, which is defined as the electric potential at contact for two charges ${q_i}$ with respect to the thermal energy ${k_{B}T}$:
\begin{equation}
    \epsilon=\frac{U_{elec}(2a)}{k_BT}=\frac{q_{i}^2}{8 \pi \epsilon_{f}a k_BT}=\frac{\lambda_{\mrm{B}}}{2a} \;,
    \label{eq:str}
\end{equation}
where $\lambda_{\mrm{B}}$ is the Bjerrum length and $a$ is the \emph{hydrated} ion radius.\cite{Israelachvili1991} This parameter, $\epsilon$, provides information on how coupled charged species are and sets the charge $q_i=\sqrt{8\pi a \varepsilon_f \epsilon k_B T}$ and charge scale ${\sqrt{a\varepsilon_f k_B T}}$. When $\epsilon$ is less than one, thermal fluctuations can easily overcome transient ion pairs. In this regime, if concentrations are also low, correlations can be ignored and the Debye length may describe interactions. As the ratio between the strengths of electrostatic forces with respect to the thermal energy increases, $\epsilon > 1 $, ions become more correlated. The colloid bead charge with respect to the ion charge is given by $\tilde{\sigma}=\frac{Q a^{2}}{4\pi a_{p}^2 \tilde{q_i}}$. Where ${Q}$ is the net charge of the colloid and the tildes correspond to non-dimensional quantities. All simulations were set to ${\tilde{\sigma}=1.0}$. Snapshots of some of the simulations are shown in figure \ref{fig:Snapshots}.\\
\indent We vary the nominal (disregarding the volume occupied by the colloidal particles and the added counterions) ion volume fraction from $\phi_i = 0.001$ to $\phi_i = 0.55$.  The corresponding molarities are highly sensitive to the hydrated radius of the ion, $a$. As an estimate $\phi_i = 0.001$ and $\phi_i = 0.55$ would correspond to approximately $3\;\mrm{mM}$ and $1.8\;\mrm{M}$ of $\text{NaCl}$ if $a\approx 0.4 \mrm{nm}$, but larger molarities are reached for ionic liquids and other macroions. We perform our analysis for two different sizes of colloidal particles, two equally sized $a_p \approx 7.5 a $ and two equally sized $a_p \approx 15 a $, which correspond to colloids of approximately $6\;\mrm{nm}$ and $12\;\mrm{nm}$ in size.
The volume fraction of the colloidal particles was set to $\phi_p=0.004$ and $\phi_p=0.008$, fixing the volume of the simulation boxes to $V=L^3$ with $L=130a$ (approximately $52\;\mrm{nm}$, assuming $a\approx 0.4$) and $L=200a$ (approximately $80\;\mrm{nm}$, assuming $a\approx 0.4$), for $a_p \approx 7.5 a $ and $a_p \approx 15 a $ respectively.\\
\indent The inherent tendency of simulations to linger in low-energy regions poses a significant challenge when computing the expectation weights for $\ave{\chi_j }$, often necessitating impractically long simulation times to thoroughly sample the entire energy landscape. To avoid this challenge, we employ a classical importance sampling technique known as ``Umbrella Sampling''\cite{torrie1977}. As discussed in sec. \ref{sec:Methods}, this technique takes advantage of the use of external biasing potentials to sample high-energy regions in the energy landscape. Figure \ref{fig:Snapshots} illustrates examples of the implementation of external biasing potentials for two negatively charged colloidal particles in a dilute electrolyte (top row) and a concentrated electrolyte (bottom row). The positive and negative species are shown in red and blue, respectively.  Subpanels a) to c) and d) to f) show snapshots of simulations biased at three different resting lengths for the dilute and concentrated regimes, respectively. The resting distance of the biasing potential decreases as we move to the right in the figure. The upper row corresponds to a nominal ion concentration of $\phi_i = 0.01$ ( approximately $30\; \mrm{mM}$ of $\text{NaCl}$ if $a\approx0.4\;\text{nm}$), and the lower row corresponds to $\phi_i = 0.20$ ( approximately $600\; \mrm{mM}$ of $\text{NaCl}$ if $a\approx0.4\;\text{nm}$). Subpanels c) and f) of Figure \ref{fig:Snapshots} are examples of high energy configurations that are unlikely to be encountered in the absence of the biasing potential. In these scenarios, the repulsive colloidal particles are positioned in close proximity, which is unlikely to occur without the presence of the externally applied biasing potential. The ability to simulate and analyze such configurations provides invaluable insights into the behavior and interactions of colloidal particles under a variety of conditions, enhancing our understanding of these complex systems. There are several established methods for reconstructing the unbiased PMF from umbrella sampling \cite{chipot_free_2023}, we employ the Multistate Bennett Acceptance Ratio (MBAR) estimator.\cite{Shirts2008} The biasing potential used in our study and the most important equations needed for estimating the underlying PMF are discussed in sec. \ref{sec:Methods}. \\
  \begin{figure}
\includegraphics[width=0.5\textwidth]{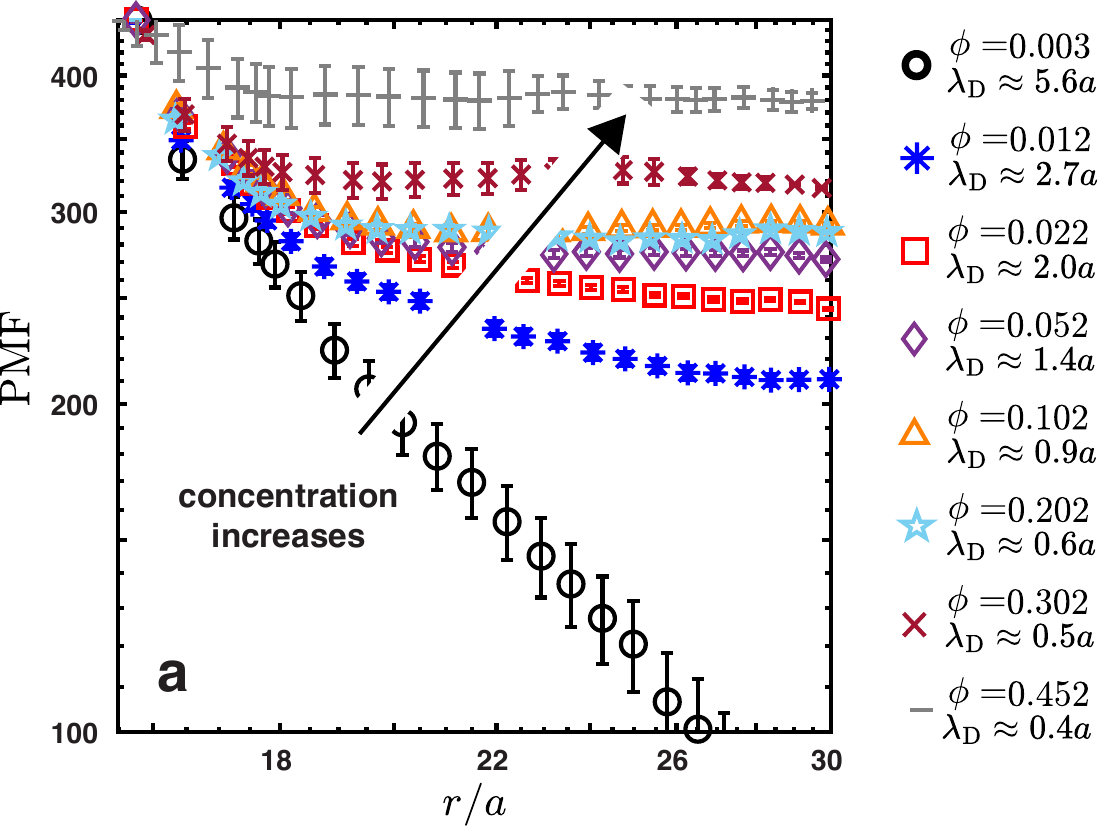}
    \caption{\label{fig:PMF} a) Potential of mean force in units of reduced energy (in units of $k_\mrm{B}T$) between two equally charged colloidal particles with radii $a_p \approx 7.5 a $.The total volume fraction occupied by all the ionic species (including added counterions and disregarding the volume occupied by the colloids) varies from $\phi = 0.003$ (circled black markers) to $\phi = 0.452$ (gray dash). For visual clarity, all energies in this figure were moved by an additive constant such that they all share the same estimated energy at the shortest sampled distance. 
    }
\end{figure}

\indent An example of the resulting PMFs is shown in Figure \ref{fig:PMF} for the case of $\epsilon = 2.0 $ and $a_p \approx 7.5a $. The $x$ axis is the center-to-center distance between the colloids in units of the hydrated ion radius, $a$. The $y$ axis is the reduced PMF with respect to the thermal energy, $k_\mrm{B}T$. The crossover of the $y$ axis was translated so that all PMFs share the same value at the smallest sampled distance. This does not alter our results because we are interested in \emph{energy differences}. The legends in Fig. \ref{fig:PMF} correspond to the total volume fraction of the ions (including the colloid's counterions). Additional PMFs are found in the SI.\\
\indent The estimated PMFs reveal a distinct behavior in systems where the screening length is shorter than the ion diameter, specifically when $\lambda_{\mathrm{D}} < 2a$ and $\phi_i < 0.30$. Under these conditions, the PMF exhibits decaying oscillations with a periodicity larger than the ionic diameter. Crucially, the origin of these oscillations differs from the well-documented structural forces arising from the displacement of discrete ion layers between colloidal particles, as observed in experiments and molecular dynamics (MD) simulations of Cd-terminated $\mathrm{CdSe}$ nanoparticles in molten $\mathrm{KCl}$ by Zhang and collaborators.\cite{zhang_stable_2017} \\
\indent Instead of a structural origin, we propose that these oscillations stem from an electrostatic mechanism, consistent with the theoretically predicted Kirkwood transition. This transition marks a shift from monotonic decay to damped oscillatory behavior in the asymptotic form of the charge-charge correlation function for bulk ions, as first described by Kirkwood.\cite{kirkwood_statistical_1936} Experimental evidence further corroborates this transition.\cite{fetisov_nanometer-scale_2020} Notably, the Kirkwood transition signifies a profound deviation from the ideal solution behavior: bulk ions transition from a state of local electroneutrality to one characterized by periodic charge density oscillations. Such oscillations, predicted even by ionic models without excluded volume interactions,\cite{Andelman2019} are supported by integral equation theories\cite{carvalho_decay_1994} and have been observed in particle-based simulations.\cite{krucker-velasquez_2021}\\

\begin{figure*}[ht]
\centering
\includegraphics[width=0.85\textwidth]{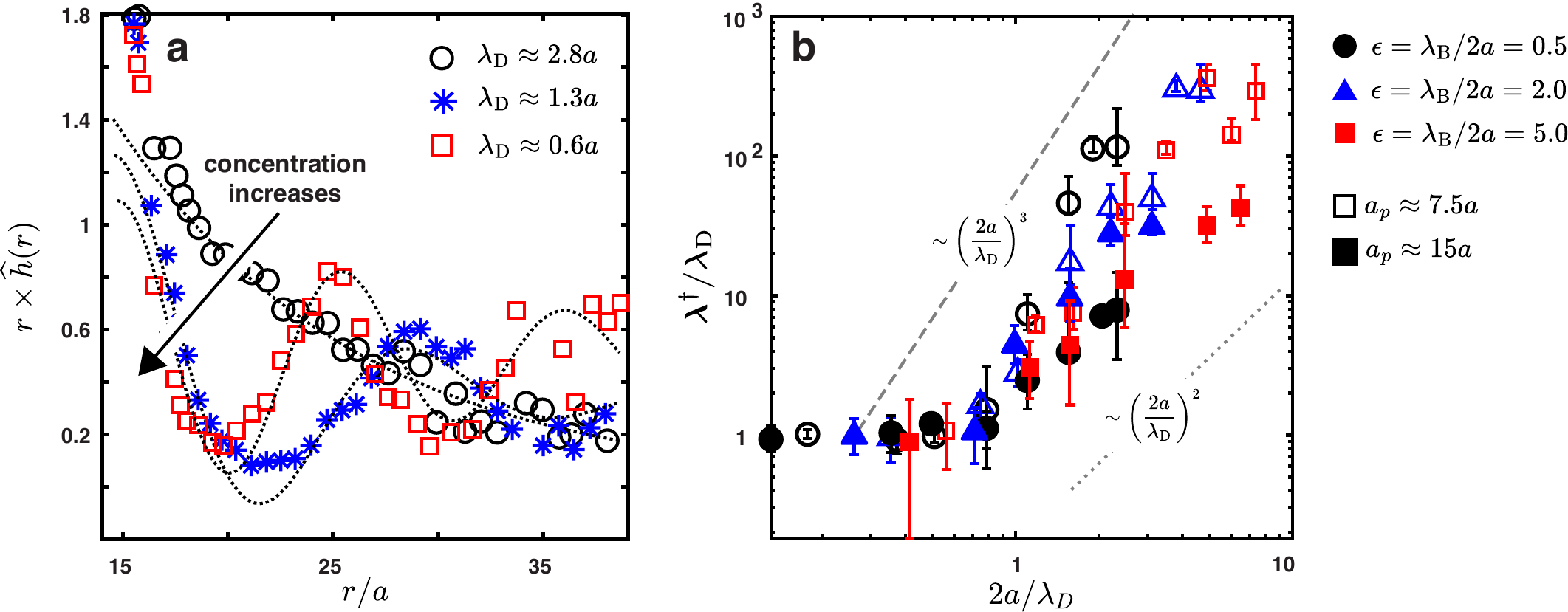}
\caption{ a) Oscillations of the total correlation function (inferred from the PMFs and not the pair distribution function ), $r\times \hat{h}(r)$ as a function of the distance between the colloidal particles. The values of s $\hat{h}(r)$ as shown here have been shifted vertically for visual clarity. Notice that for the lowest concentration ($\lambda_{\mathrm{D}}\approx 2.8 a$), oscillations of the correlation function have not been fully developed. In contrast, $\lambda_{\mathrm{D}}\approx 1.3 a$, and $\lambda_{\mathrm{D}}\approx 0.6 a$ show fully developed oscillations of the total correlation function. The dashed lines correspond to the best fit using a single mode from equation \ref{eq:hofr}. The computed $\lambda_\mathrm{D}$ accounts for the added counter ions and disregards the volume occupied by the colloids.  b) Estimated decay length, $\lambda^{ \dagger }$, in units of the Debye screening length, $\lambda_{\mrm{D}}$, as a function of the inverse Debye screening length in units of the diameter of the hydrated ion, $2a$. The dotted and dashed gray lines indicate scalings for $n=2$ and $n=3$, respectively, in $\left( \lambda^{\dagger} / \lambda_{\mrm{D}} \right)\sim (2a/\lambda_{\mrm{D}})^n$.}\label{fig:scaling}
\end{figure*}
\indent As the potential of \emph{mean} force is inherently an integrated quantity, these oscillations are best brought to the fore via the total correlation function,$h(r)=g(r)-1$. If adhering to the principle of maximum work, then $\mrm{PMF}(r) = - k_{\mrm{B}}T \log g(r)$. Traditionally, the correlation function $h(r)$ is derived from the pair distribution function $g(r)$, rather than directly from the forces that act on the particles. However, by adopting this concept from liquid state theory, we can more effectively discuss the oscillations observed in the potential of mean force (PMF). To clarify this distinction, we will denote the correlation function obtained from the PMF, rather than $g(r)$, by $h(r)$.  Figure \ref{fig:scaling} a) shows $r\times \hat{h}(r)$ as a function of the distance between colloidal particles in our simulations, where the hat indicates that this quantity has been rescaled by its value at the shortest sampled distance for easier visualization. The absence of charge oscillations at a nominal concentration (not taking into account the colloidal particles' counter-ions) of $\phi = 0.05$ (black circled markers) starkly contrasts with the pronounced oscillations observed at concentrations exceeding the Kirkwood transition, as seen at $\phi = 0.10$ (starred blue markers) and $\phi = 0.30$ (squared red markers). \\
\indent Analytical solutions for $h(r)$ commonly take the form of an infinite sum over relaxation modes:\cite{hansen_mcdonald_2006}  
\begin{equation}
    h(r)\sim \sum_j \left(cos(\omega_j r + \psi_j) \right)\frac{\exp(-r/\lambda_j) }{r} \; ;
    \label{eq:hofr}
\end{equation}
with $\omega_j$ and $\psi_j$ representing the frequency and phase of the $j^{\mrm{th}}$ relaxation mode of charge density oscillations, respectively. 
In the absence of oscillations, $\omega_j = 0$, the modes are purely exponential. When $\omega_j > 0$ these oscillations decay exponentially, governed by the $j^{\mrm{th}}$ correlation length. Although there can be multiple correlation lengths,\cite{kjellander_intimate_2019} we are interested in the far-field correlation length given by a single mode in eq. \ref{eq:hofr} that endures as $r\rightarrow\infty$.\cite{Andelman2019} The prevailing correlation length in the far field is $\lambda_\mrm{D}$ at low ion concentrations with $\omega_j = \psi_j = 0$. \\
\indent We note fitting the oscillations is far from trivial when inferred from the PMFs due to the model's high nonlinearity near $\omega_j \approx 0$ and the inferred PMF's inability to produce numerically tractable oscillations. Nonetheless, our previous work\cite{krucker-velasquez_2021} computed such oscillations for \emph{ions in the bulk} using spectral methods that are better equipped to accurately extract long-range structural features. Moreover, we consistently found truncating eq. \ref{eq:hofr} to the first relaxation mode to be sufficient to capture the far-field properties of the suspension. Hence, we opt to extract the correlation length by fitting only the asymptotic expression $ \exp(-r/\lambda_{\mrm{D}})/r $ to the PMF. \\
\indent The estimated correlation lengths obtained from the PMFs are summarized in Figure \ref{fig:scaling} as a function of the Debye screening length in units of the diameter of the hydrated ion. Clear markers indicate correlation lengths obtained from simulations of colloidal particle radii of $a_p \approx 7.5 a$, and filled markers of colloidal particles $a_p \approx 15 a$.  The dotted and dashed gray lines show two different scalings for the estimated correlation lengths in units of the Debye screening length, $\left( \lambda^{\dagger} / \lambda_{\mrm{D}} \right)$. The dotted line shows the $\left( \lambda^{\dagger} / \lambda_{\mrm{D}} \right)\sim (2a/\lambda_{\mrm{D}})^2$ scaling, and the dashed line shows the $\left( \lambda^{\dagger} / \lambda_{\mrm{D}} \right)\sim (2a/\lambda_{\mrm{D}})^3$ scaling. Most of the values where $\left( \lambda^{\dagger} / \lambda_{\mrm{D}} \right) \geq 1 $ follow superlinear scalings where $n \geq 2.5$ and $n \leq 3.2$. \\
\indent An interesting trend arises in Fig. \ref{fig:scaling} when $2a/\lambda_{\mrm{D}}\geq1$. In particular, the ratio $\lambda^{\dagger}/\lambda_{\mrm{D}}$ appears to depend not only on the radius of the ion diameter, but also on the radius of the colloidal particles. This can be observed by noticing that the values obtained for the largest colloid size (filled markers) fall consistently below the values for $a_p\approx7.5a$ (clear markers). This trend suggests that the size of the colloidal particles might control the highly concentrated regime. This is consistent with the introduction of a new correlation length $\lambda^{\dagger \dagger}$ in units of $a_p$ that satisfies $\lambda^{\dagger}\sim \lambda^{\dagger \dagger}/a_p $.\\
\section{Conclusions}

\indent Our investigation of the potential of mean force (PMF) between metallic nanoparticles expands the current understanding of interparticle forces in electrostatically stabilized colloidal suspensions. At low ionic concentrations, charged metallic particles exhibit short-range oscillatory forces at small separations, followed by a monotonic decay at larger distances, consistent with extensive experimental validation.\cite{Israelachvili1991} However, at higher ionic concentrations, where $\lambda_{\mathrm{D}} < 2a$, the PMF exhibits nonmonotonic decay at long distances. Two critical distinctions emerge between the low- and high-concentration regimes: the presence of long-range oscillations in the PMF that lack a structural origin and a significantly slower decay of interparticle forces than predicted by DLVO theory.\\

\indent These oscillations, characterized by wavelengths larger than the ionic diameter for concentrations below $\phi_i < 0.30$, suggest a fundamentally different origin from the short-range structural forces associated with the packing of solvent layers or other species. The distinct length scale, extended range, and alignment of these oscillations with the Kirkwood transition strongly support their electrostatic nature. Although subtle, these oscillations may induce metastable states in colloidal suspensions, supplementing those governed by the primary and secondary energy minima predicted by the DLVO theory.\\
\indent The rate of decay of interparticle forces in concentrated electrolytes, governed by the effective screening length $\lambda^{\dagger}$, diverges from that observed in dilute electrolytes, where it aligns well with $\lambda_{\mathrm{D}}$. When $\lambda_{\mathrm{D}} < 2a$, ionic screening appears to suppress electrostatic interactions less effectively than in the dilute regime. Notably, we identify a range of scaling exponents for the relation $\left( \lambda^{\dagger} / \lambda_{\mathrm{D}} \right)\sim (2a/\lambda_{\mathrm{D}})^n$, bounded between $n=2.3$ and $n=3.2$, with most data points clustering between $n=2.5$ and $n=3.2$. While these scaling regimes offer a useful framework for characterizing interparticle potentials and ionic structure, they should be regarded as approximations rather than absolute descriptors when applied to engineering applications.\\

\indent Although the effective screening lengths in this study were calculated in the presence of metallic surfaces, our findings share similarities with anomalous underscreening. A plausible explanation for this resemblance lies in the additional geometric constraints imposed by charged metallic surfaces. Charged surfaces profoundly influence the spatial arrangement of ionic species nestled between colloidal particles. The presence of these charged interfaces initiates charge oscillations that extend outward, shaping the long-range structure of the ions in a manner distinct from the free rearrangement expected within the bulk environment. This modulation of the ionic structure by charged surfaces underscores a pivotal shift from bulk behavior, where ions navigate and position themselves without the directive influence of external charged boundaries. Consequently, this interaction mechanism plays a crucial role in defining the electrostatic potential landscape of the system, guiding the distribution and organization of ions in the vicinity of colloidal entities and altering the fundamental nature of ionic interactions in these confined spaces. Notably, as Fig. \ref{fig:scaling} suggests, there is a dependence on the size of the colloidal particles when $2a/\lambda_{\mrm{D}}>1$, implying that this length scale could be pivotal in controlling electrostatic forces in highly concentrated regimes.\\
\indent By investigating how efficiently electrolytes dampen electrostatic interactions in metallic colloidal particles, considering variations in ionic concentration, charge-charge interaction strength (analogous to temperature changes), and colloidal particle size, we seek to elucidate the mechanisms behind deviations from classical Debye-Hückel (DH) theory in concentrated electrolytes, such as underscreening and other anomalous behaviors observed in charged soft matter systems.\cite{reinertsen_reexpansion_2024,li_salt-dependent_2023} This exploration not only advances our understanding of colloidal systems across diverse environments but also provides valuable scaling trends for effective screening lengths, serving as a guide for designing engineering applications in nanotechnology.\\
\section{\label{sec:Methods}Methods}
\subsection{Brownian Dynamics}
\indent We remind the reader that we are simulating two colloidal particles in an explicit electrolyte. In our study, all charged species are simulated explicitly, and the solvent is treated implicitly. At the length scales of interest for our study, diffusion relaxation time is significantly larger than the inertial relaxation time of the particles, allowing us to neglect inertial effects. The species' dynamics are thus governed by the overdamped Langevin equation:\cite{Frenkel2002}
\begin{equation}
\mathbf{F}_{\alpha}^{\mathrm{H}}+\mathbf{F}_{\alpha}^{\mathrm{I}}+\mathbf{F}_{\alpha}^{\mathrm{E}}+\mathbf{F}_{\alpha}^{\mathrm{B}}=0\:,
\label{eq:overdamped}
\end{equation}
where $\mathbf{F}_{\alpha}^{\mathrm{H}}$ denotes the hydrodynamic force, $\mathbf{F}_{\alpha}^{\mathrm{I}}$ the forces from a conservative potential, $\mathbf{F}_{\alpha}^{\mathrm{E}}$ the external force, and $\mathbf{F}_{\alpha}^{\mathrm{B}}$ the stochastic Brownian force. This last force complies with the fluctuation-dissipation theorem\cite{Russel1989}.\\
\indent The hydrodynamic mobility tensor, $\mathbf{M}^{\mathrm{H}}$, links the non-hydrodynamic forces of the $\beta^{\mrm{th}}$ particle, $\mathbf{F}_{\beta}=\mathbf{F}_{\beta}^{\mathrm{I}}+\mathbf{F}_{\beta}^{\mathrm{E}}+\mathbf{F}_{\beta}^{\mathrm{B}}$, to the velocities of the $\alpha^{\mrm{th}}$ particle through $\mathbf{u}_{\alpha}(t)= \sum_{\beta=1}^{N}\mathbf{M}_{\alpha \beta}^{\mathrm{H}}\cdot\mathbf{F}_{\beta}(t)$.Equation \ref{eq:overdamped} is numerically solved via a forward Euler-Maruyama integration scheme.\cite{Swan2016,Fiore2017} Considering the equilibrium nature of the PMF and the dissipative nature of interparticle hydrodynamic forces, we simplify the model by setting the drag equal to the Stokes drag for each bead and ignore all interparticle hydrodynamic interactions.\\
\indent The forces from conservative interactions among ions are described by the gradient of a potential energy $U(\mathcal{X})$, which depends on the coordinates of all ions $\mathcal{X} \equiv [\mathbf{x}_1, \mathbf{x}_2, \dots, \mathbf{x}_N]^T$:
\begin{equation}
 \mathbf{F}_{\alpha}^{\mathrm{I/E}}(\mathcal{X}) \equiv -\nabla_{\mathbf{x}_{\alpha}} U^{\mathrm{I/E}}(\mathcal{X}),   
\end{equation}
where the gradient is taken with respect to the position of the $\alpha$\textsuperscript{th} particle. To prevent particle overlap, we employ the Heyes and Melrose hard-sphere algorithm\cite{Heyes1993, Heyes1994A}, ensuring the physical integrity of the simulated charge system.\\
\indent We obtain the electrostatic forces, $\mbf{F}^{\mrm{E}}$, for the ionic and charged colloids by solving Poisson's equation for the electrostatic potential $\psi(\mathbf{x})$ on the particles and on the fluid. If we assume all particles are immersed in a constant dielectric and all the charge is localized on the surface of the ions and/or the colloidal particles, Poisson's equation can be simplified to Laplace's equation,
\begin{equation}
    \nabla^2\psi=0\:,
    \label{eq:laplace}
\end{equation}
subject to the boundary conditions on the surfaces of the charged species:
\begin{equation}
    \psi_{\mathrm{p}} = \psi_{\mathrm{f}} , \hspace{4mm} (\varepsilon_\mrm{f} \mathbf{E}_{\mathrm{f}} - \varepsilon_\mrm{p} \mathbf{E}_{\mathrm{p}}) \cdot \hat{\mathbf{n}} = q_{\alpha}/(4 \pi a^2)\:, \hspace{2mm}
    \label{eq:BCs}
\end{equation}
where ${\psi_{\mathrm{p}}}$ and ${\psi_{\mathrm{f}}}$ are the potentials inside and outside the particle, ${q_{\alpha}}/(4\pi a^2)$ is the uniform free surface charge density of ion $ \alpha $ on a spherical shell with radius $ a $ and net charge ${q_{\alpha}}$,  $\mathbf{\hat{n}}$ is the normal outward vector and ${\mathbf{E}_{\mathrm{f}} }$ and ${\mathbf{E}_{\mathrm{p}} }$ are the electric field outside and inside the particle, respectively. \cite{Jackson1999} 
We assume the time scale prescribed by particle motion is significantly slower than the electric and magnetic relaxation at the atomic scale, and consequently the time dependence of the electric potential emerges solely through the time-varying boundary conditions in eq. \eqref{eq:BCs}.\\
\indent Starting from the potential given by the integral form of Laplace's equation, as described in detail in ref. \cite{krucker-velasquez_immersed_2024}, we can express the electric potential for a point ${\mathbf{x}}$ in the fluid in terms of multipole moments via a spherical harmonic expansion,
\begin{align}
    &\psi_{\mathrm{f}}(\mathbf{x})-\psi_{0}(\mathbf{x})=\\ \nonumber
    &\frac{\varepsilon_{\mathrm{f}}}{V}\sum_{\mathbf{k}\neq0}\sum_{\alpha}\frac{e^{i\mathbf{k}\cdot(\mathbf{x}-\mathbf{x}_{\alpha})}}{k^{2}} \left( q_{\alpha}j_{0}(ka) - \frac{ 3 i }{ a } j_{1}(ka) \mbf{S}_{\alpha} \cdot \hat{ \mbf{k} } \cdots \right) \;,
    \label{eq:psi_f}
\end{align}
where ${j_{i}(x)}$ is the spherical Bessel function of the first kind of $i^{\mrm{th}}$ order, $\mathbf{k} = [2\pi \kappa_x/L_x, 2\pi \kappa_y/L_y, 2\pi \kappa_z/L_z]$ for integers $\kappa_i$, and $\hat{\mbf{k}}$ the unit vector in the $\mbf{k}$ direction.\\
\indent Similarly, we can represent the potential on the particles as a local expansion about the $\beta^{\mrm{th}}$ particle. We integrate on the surface of each particle applying the boundary conditions in eq. \eqref{eq:BCs}, and retaining only the first moment of expansion such that
\begin{equation}
\psi_\alpha(\mathbf{x}_\alpha) - \psi_0(\mathbf{x}_\alpha) = \frac{\varepsilon_{\mathrm{f}}}{V}\sum_{\mathbf{k}\neq0} \sum_\beta j_{0}^{2}(ka)\frac{e^{i\mathbf{k}\cdot(\mathbf{x}_{\alpha}-\mathbf{x}_{\beta})}}{k^{2}} q_\beta \; .
\label{eq:pot_charge}
\end{equation}
where $\psi_\alpha = \frac{1}{4 \pi a^2 } \int_{S_{\alpha}} \mrm{d} \mbf{x}  \psi_{\mrm{p}}(\mbf{x})$ is the surface average $\psi_\mrm{p}$ of the particle $\alpha$. By defining the elements of the potential matrix as $M^{\mrm{E}}_{\alpha\beta} = \frac{\varepsilon_{\mathrm{f}}}{V}\sum_{\mathbf{k}\neq0}j_{0}^{2}(ka)\frac{e^{i\mathbf{k}\cdot(\mathbf{x}_{\alpha}-\mathbf{x}_{\beta})}}{k^{2}}$, the system of non-linear equations can be shown in the matrix-vector form given by
\begin{equation} \label{eq:Moments}
\Psi -\Psi_{0}=\mathbf{M}^{\mrm{E}}\cdot\mathcal{Q}\;,
\end{equation}
where $\Psi -\Psi_{0}=[\psi_{1} -\psi_{0}(\mathbf{x}_{1}), \psi_{2} -\psi_{0}(\mathbf{x}_{2}),...,\psi_{N} -\psi_{0}(\mathbf{x}_{N})]^{T}$
is a list of particle potentials relative to the external potential at their centers, $\mathcal{Q} =[q_{1},q_{2},...,q_{N}]^T$ is the list of particle charges. This system of equations relates each of the $N$ particle potentials $\psi_\alpha$ to the $N$ particle charges $q_\beta$. We use the general minimized residual (GMRES) method \cite{freund_iterative_1992} to solve this system of equations in combination with the Spectral Ewald method described in detail in refs. \cite{krucker-velasquez_immersed_2024} and \cite{Lindbo2011}.\\
\indent The total electric potential energy of the electrolyte is:
\begin{equation}
    U^{E}= \frac{1}{2}\mathcal{Q}\cdot \Psi = \frac{1}{2}\mathcal{Q}\cdot \Psi_0 + \frac{1}{2}\mathcal{Q}\cdot( \Psi -\Psi_{0}).
    \label{eq:E_potential}
\end{equation}
Taking the gradient of the electric potential energy, we can calculate the electric force on the ${\alpha^{\mathrm{th}}}$ particle:
\begin{equation}
    \mathbf{F}_{\alpha}^{\mathrm{E}}=-\sum_{\beta}\nabla_{\mathbf{x}_{\beta}}\mbf{M}_{\psi q}^{E}q_{\alpha}q_{\beta}.
    \label{eq:electric_force}
\end{equation}
This force can be computed using the same matrix free methods explained in ref. \cite{krucker-velasquez_immersed_2024} as the potential.\\
\subsection{ Induced Surface Charge Distribution on the Metallic Colloids}
\indent For a suspension of $N_b=2$ colloids, where the center of mass of the ${i^\mathrm{th}}$ colloid is $\mathbf{X}_{i}$. The surface of each colloid is discretized with $N_{i}$ beads of radius $a$ at positions $\mathbf{x}_{i j}$. In our simulations, we considered two different particle sizes. For $a_b\approx7.5$ and $a_b\approx15$, $N_i=162$ and $N_i=642$, respectively. Ionic species can be considered to be colloids with one particle. Lowercase bead quantities with two indices ($i j$) indicate the $j^\text{th}$ bead on the $i^\text{th}$ colloid, while a single index ($\beta$) is instead a global bead index (\emph{i.e.} $\beta = (i-1)N_i + j$). First, we consider charge conservation in each of the colloids, consequently, the total charge on the colloid ${i^\mathrm{th}}$ will be equal to the sum of the charges of its constituent beads $q_{i j}$, $Q_{i} = \sum_{j} q_{i j}$.
In metallic particles, each of the constituent bead's potential equals the colloid's potential, $\psi_{i j } = \Psi_{i}$ such that difference between the bead potential and the external potential is
\begin{equation} \label{eq:beadpot1}
\psi_{i j } - \psi_{0,i j } = \Psi_{i} - \Psi_{0,i} + \mathbf{r}_{i j } \cdot \mathbf{E}_0,
\end{equation} 
where $\mathbf{r}_{i j} \equiv \mathbf{x}_{i j} - \mathbf{X}_{i}$ is the position of the bead relative to the center of mass of the particle. The bead potentials $\phi_\alpha$ are also related to the body charges $q_\beta$ through the potential tensor $\mathbf{M}^{E}$ in eq. \eqref{eq:pot_charge}. We can construct a system of equations for the charge distribution by combining charge conservation on each colloid and eq. \eqref{eq:beadpot1}
\begin{equation} \label{eq:saddle}
\begin{bmatrix}
-\mathbf{M}^{\mrm{E}} & \boldsymbol{\Sigma}^T \\
\boldsymbol{\Sigma} & 0
\end{bmatrix} \cdot \begin{bmatrix}
\mathbf{q} \\
\boldsymbol{\Psi} - \boldsymbol{\Psi}^0
\end{bmatrix} = \begin{bmatrix}
-\mathbf{r} \cdot \mathbf{E}^0 \\
\mathbf{Q}
\end{bmatrix}.
\end{equation}
$\mathbf{q} \equiv [q_{11}, q_{12}, \cdots, q_{21}, q_{22}, \cdots]^T$ is a list of all $N$ bead charges, $\mathbf{Q} \equiv [Q_1, Q_2, \cdots]^T$ is a list of all $N_b$ body charges, $\boldsymbol{\Psi}-\boldsymbol{\Psi}_0 \equiv [\Psi_1-\Psi_{0,1}, \Psi_2-\Psi_{0,2}, \cdots]^T$ is a list of the difference between the $N_b$ body potentials and the external potential, and $\mathbf{r} \cdot \mathbf{E}_0 \equiv [\mathbf{r}_{11}\cdot \mathbf{E}_0, \mathbf{r}_{12}\cdot \mathbf{E}_0, \cdots, \mathbf{r}_{21}\cdot \mathbf{E}_0, \mathbf{r}_{22}\cdot \mathbf{E}_0, \cdots]^T$ is a list of $N$ relative bead positions dotted with the external field. $\boldsymbol{\Sigma}$ is an $N_b\times N$ summation tensor whose rows correspond to colloidal bodies and columns correspond to beads. Each row is entirely $0$, except for $N_{i}$ consecutive $1$ values in the columns corresponding to the $N_{i}$ beads of the $i^{\mrm{th}}$ body. Further details for constructing $\boldsymbol{\Sigma}$ and subsequent summation tensors are found in the Appendix of ref. \cite{krucker-velasquez_immersed_2024}. Eq. \ref{eq:saddle} can be solved iteratively for the induced charge and rigid body potentials. \\
\indent The induced dipole moment of a colloid can then be computed from the bead charges $\mathbf{S}_{i} = \sum_{j} \mathbf{r}_{ i j } q_{ i j }$. The set of all rigid body moments (charges and dipoles) is then given by
\begin{equation}
\begin{bmatrix}
\mathbf{Q} \\
\mathbf{S}
\end{bmatrix} = \boldsymbol{\Sigma}' \cdot \left(\mathbf{M}^{\mrm{E}}\right)^{-1} \cdot \boldsymbol{\Sigma}'^T \cdot \begin{bmatrix}
\boldsymbol{\Psi}-\boldsymbol{\Psi}_0 \\
\mathbf{E}_0
\end{bmatrix}.
\label{eq:QS_capacitance}
\end{equation}
where $\boldsymbol{\Sigma}'$ is an $4N_b\times N$ summation tensor, whose first $N_b$ rows are identical to $\boldsymbol{\Sigma}$ and whose next $3N_b$ rows look structurally like $\boldsymbol{\Sigma}$ but with each $1$ value replaced with $\mathbf{r}_{ij}$ (see Appendix in ref. \cite{krucker-velasquez_immersed_2024}).
\indent The potential energy of the suspension is expressed as the sum of products of rigid body moments and potential derivatives \cite{Jackson1999}
\begin{equation}
    U =\frac{1}{2}\mbf{Q}\cdot\boldsymbol{\Psi}_{0} + \frac{1}{2} \left[ \mathbf{q} \quad \left\langle \boldsymbol{\Psi}\right\rangle -\boldsymbol{\Psi}_{0} \right] \cdot \begin{bmatrix}
-\mathbf{r} \cdot \mathbf{E}_0 \\
\mathbf{Q} 
\end{bmatrix}\;.
\end{equation}
\indent The force on each bead is the negative derivative of $U$ with respect to the bead position $\mathbf{x}_\alpha$, given all other beads remain fixed, $\mbf{f}_\alpha = -\nabla_{\mathbf{x}_\alpha} U$. As derived in ref. \cite{krucker-velasquez_immersed_2024} the force on bead $\alpha$ is
$\mathbf{f}_\alpha = q_\alpha \mathbf{E}_0 - \frac{1}{2} \left( \nabla_{\mathbf{x}_\alpha} \mathbf{M}^{\mrm{E}} \right) : \mathbf{q}\mathbf{q} $. From the distribution of bead forces, we can compute the $i^{\mrm{th}}$ body's force, $\mathbf{F}_{i} = \sum_{j} \mathbf{f}_{i j }$ and torque $\mathbf{L}_{i}  = \sum_{j} \mathbf{r}_{i j } \times \mathbf{f}_{i j }$. We note that for ionic species, $\mathbf{r}_{ij}=0$.

\subsection{Umbrella sampling and Multi-state Bennet Acceptance Ratio estimation of the potential of mean-force}

\indent The hallmark of Umbrella Sampling is the implementation of external biasing potentials, typically spring potentials (the ``umbrella'')  of the form $U^{\mrm{B}}_k(r)= \left(k_{s,k}/2\right)\left(r-r_{0,k}\right)^2 $. The biasing potentials are strategically chosen to constrain the center-to-center distance explored from the colloids (our reaction coordinate),$r$, to a specific region of the thermodynamic energy landscape established by $r_{0,k}$. The total potential energy in each state is the sum of the true underlying potential energy, $U(r)$, and the biasing potential, $U^{\mrm{B}}_k(r)$, resulting in a biased potential energy $U^\star_k(r)=U^{\mrm{B}}_k(r) + U(r) $.  \\
\indent  Each umbrella will produce unique expectation values for some mechanical observable $A(x)$. This observable in our studies is $\chi_j(r)$. For large datasets, the expectation of $A(x)$ at the $k^\mrm{th}$ state will be defined by the probability density function associated with the $k^\mrm{th}$ state, $p_k(x)$.\cite{Macquarrie2000} That is, $\ave{A}_k = \int A(x) p_k(x) \mrm{d}x $. In equilibrium thermodynamics, the probability $p_k(x)$ is related to the the Boltzmann factor,$q_k(x)=\exp(-U_{k}^{\mrm{B}}/ k_\mrm{B} T )$, via $p_k(x)=Z_k^{-1} \exp(-U_{k}^{\mrm{B}}/ k_\mrm{B}  T )$, where $ Z_k= \int q_k(\mbf{x})dx$ is the partition constant.\cite{Macquarrie2000,Shirts2008}\\
\indent The biasing potential was constructed by connecting the colloidal particles via two co-linear spring potentials. A diagram depicting the co-linear spring potentials is present in Figure 1 a) in the SI. The total biasing potential for each thermodynamic state is 
\begin{align}
      U^{\mrm{B}}_{k}(r) & =U^{\mrm{B}}_{k,1}(r) + U^{\mrm{B}}_{k,2}(r) \nonumber \\
      &=\frac{{k_s}_{k}}{2}\left( (r_{k,1}^{0}-r_{k,1})^2 + (r_{k,2}^{0}-r_{k,2})^2 \right) \;,
  \label{eq:SIBiasing}
\end{align}
where $r_{1,k}^{0}$ is the $k^\mrm{th}$ spring's resting distance for the first spring potential ( $A-D$ spring potential in fig. 1 a) found in the SI ) and $r_{k,2}^{0}=r_{k,1}^{0}+4a_p$ ( $B-C$ spring potential in fig. 1 a) found in the SI). 
\subsection{ Multi-state Bennet Acceptance Ratio estimation of the potential of mean force}
Although the expectation values of $A(x)$ are specific to the biased states, they encompass the necessary information to reconstruct an unbiased PMF. There are several established methods for reconstructing the unbiased PMF from the configurations sampled using umbrella sampling; some of these are the Weighted Histogram Analysis Method (WHAM)\cite{kumar_weighted_1992}, Bennett Acceptance Ratio (BAR)\cite{bennett_efficient_1976}, and Thermodynamic Integration\cite{shirts_comparison_2005,chipot_free_2023}. These techniques have been thoroughly discussed in various studies, including a recent review found in ref. \cite{chipot_free_2023}.\\
\indent We employ an extension of the BAR method known as the Multistate Bennett Acceptance Ratio (MBAR) estimator.\cite{Shirts2008} This method relates measurements between thermodynamic multi-states to make efficient use of all the sampled data. In particular, this method obtains more accurate estimates with fewer data points than BAR and the distribution of data determines uncertainties associated with these expectation values. MBAR relates the expectations sampled from the $k^\mrm{th}$ and $l^\mrm{th}$ probability distributions through a bridge function, $\alpha_{kl}$,
\begin{equation}
    z_{k}\left\langle \alpha_{kl}q_{l}\right\rangle_{k}=z_{l}\left\langle \alpha_{kl}q_{l}\right\rangle _{l}\;.
    \label{eq:bridge}
\end{equation}
Equation \ref{eq:bridge} leads to family of equations parameterized by
$\alpha_{kl}=N_{l}\hat{z}_{l}^{-1}/\sum_{m}^{M}N_{m}\hat{z}_{m}^{-1}q_{m}(\mbf{x})$,\cite{Shirts2008} where $N_{l}$ and $N_{m}$ represent the number of samples from the $l^\mrm{th}$ and $m^\mrm{th}$ thermodynamic state, respectively. Solving this system of nonlinear equations self-consistently allows us to estimate the ratios $ \hat{z}_k / \hat{z}_l$, leading to a matrix of weights whose element $nl^{ \mrm{th} }$ is
\begin{equation}
    W_{nl}=\hat{z}_l^{-1} \frac{q_l(\mbf{x}_n)}{\sum_{k=1}^{K}N_k z_k^{-1}q_{k}(\mbf{x}_n)}\;,
    \label{eq:weights}
\end{equation}
where the index $n$ runs over all sampled trajectories of the $l^{\mrm{th}}$ thermodynamic state. This is a weight matrix that satisfies $\mbf{1}\cdot \mbf{W}=\mbf{1}$ and $ \mbf{W}\cdot \tilde{\mbf{N}} = \mbf{1} $, where $\mbf{1}$ is a vector of one and $\tilde{\mbf{N}}$ a vector containing the number of samples from each thermodynamic state. \\ 
\indent To estimate the true value of $A$ (or $\chi_j$), we introduce new probability distributions $q_A = A(x) q(x) $ and $q_a = q(x)$. As the samples in states $A$ and $a$ are zero, we can use the same bridge samplers computed from the thermodynamic states sampled because they do not contribute to the sum in the denominator of $\alpha_{kl}$. Thus, we can estimate $A$ as
\begin{equation}
    \hat{A}=\frac{\hat{z}_A}{\hat{z}_a} = \sum_{n=1}^{\tilde{N}} 
 W_{na}A(x_n)\;.
\end{equation}
where $\hat{z}_A=\sum_{n=1}A(x)q_{n}(\mathbf{x}_{n})/\sum_{k=1}^{K}\tilde{N}_{k}z_{k}^{-1}q_{k}(\mathbf{x}_{k})$ and $\hat{z}_a = \sum_{n=1}q_{n}(\mathbf{x}_{n})/\sum_{k=1}^{K}\tilde{N}_{k}z_{k}^{-1}q_{k}(\mathbf{x}_{k})$. Assuming the weights are normally distributed, we can estimate the uncertainty as $ \delta^2 \hat{A} =\mrm{cov}(\frac{\hat{z}_A}{\hat{z}_a},\frac{\hat{z}_A}{\hat{z}_a}) $. 

\begin{acknowledgement}
 We acknowledge the support from NASA, Grant No. 80NSSC18K0162 and NSF, Career Award No. 1554398.
\end{acknowledgement}

\begin{suppinfo}
Details on the numerical techniques and additional potentials of mean force can be found in the Supporting Information
\end{suppinfo}

\bibliography{achemso-demo}

\end{document}